Self-Consistent and Environment-Dependent Hamiltonians for Materials Simulations:  Case Studies on Silicon Structures

C. Leahy, M. Yu, C.S. Jayanthi, and S.Y. Wu

*Department of Physics, University of Louisville, Louisville, KY 40292*

Abstract

A reliable semi-empirical Hamiltonian for materials simulations must allow electron screening and charge redistribution effects. Using the framework of linear combination of atomic orbitals (LCAO), a self-consistent and environment-dependent (SCED) Hamiltonian has been constructed for quantum mechanics based simulations of materials. This Hamiltonian contains environment-dependent multi-center interaction terms and electron-electron correlation terms that allow electron screening and charge-redistribution effects. As a case study, we have developed the SCED/LCAO Hamiltonian for silicon. The robustness of this Hamiltonian is demonstrated by scrutinizing a variety of different structures of silicon. In particular, we have studied the following: (i) the bulk phase diagrams of silicon, (ii) the structure of an intermediate-size $Si_{71}$ cluster, (iii) the reconstruction of Si(100) surface, and (iv) the energy landscape for a silicon monomer adsorbed on the reconstructed Si(111)-7x7 surface. The success of the silicon SCED/LCAO Hamiltonian in the above applications, where silicon exists in a variety of different co-ordinations, is a testament to the predictive power of the scheme.

PACS number(s): 71.15-m, 71.15.Pd, 68.35.Bs, 68.43.-h, 73.22.-f

**I. Introduction**

The premise dictating the present development of a recipe for constructing a semi-empirical, self-consistent, and environment-dependent (SCED) Hamiltonian for materials is to find a reliable and efficient scheme to mimic the effect of screening by electrons when atoms are brought together to form a stable aggregate. The goal is to construct a Hamiltonian that is transferable, and hence has predictive power. While density functional theory (DFT)-based molecular dynamics (MD) schemes for the determination of structural properties of materials are expected to have predictive power, their applications are, however, still limited to systems of about a few hundreds of atoms. Tight-binding (TB) MD schemes, on the other hand, are fast and applicable to larger systems. However, the transferability of conventional TB Hamiltonians is limited because they include only two-center interactions and they have no framework to allow the self-consistent determination of the charge re-distribution. Hence they do not have predictive power and can only be used, in the strictest sense, to provide explanation for system-specific experimental results. In recent years, various schemes have been proposed to improve the transferability of TB Hamiltonians by including the self-consistency and/or the environment-dependency[1-11]. Among these methods are those that can also be conveniently implemented in MD schemes because the atomic forces can be readily calculated for those methods. They include methods whose emphasis is placed on a phenomenological description of the environment-dependency[2,3] and two methods that are similar and whose frameworks take into account the self-consistency as well as the environment-dependency[8,11]. For example, the approach of Ref. 11 is based on the expansion of the DFT-total energy in terms of the charge density fluctuations about some reference density. To the second order in the density fluctuations[12], the total energy is approximated as the sum of a band structure term and a short-range repulsive term corresponding to the conventional two-center TB Hamiltonian, plus a term representing the Coulomb interaction between charge fluctuations. Within this framework, the charge fluctuations can be self-consistently determined by solving an eigenvalue equation with the two-center Hamiltonian modified by a term that depends on the charge redistribution. While the Hamiltonian so defined does contain the features of self-consistency in the charge redistribution and the environment-dependency for systems with charge fluctuations, the environment-dependent feature disappears when systems under consideration do not involve charge fluctuations, e.g., periodic extended systems. But the environment-dependency is a key feature in a realistic modeling of the screening effect of the electrons in an aggregate of atoms, including extended periodic systems. This deficiency in properly mimicking the screening of the electrons can be critical in the development of a truly transferable Hamiltonian. Furthermore, the construction of the Hamiltonian and the determination of the total energy are, on the one hand, dependent on an optimal basis set of confined atomic orbitals obtained by solving a modified Schrödinger equation for a free atom in the framework of a self-consistent local density approximation with the correction of the generalized gradient approximation (SC-LDA/GGA). In this way, the Hamiltonian and overlap matrix elements are determined as functions of the distance between pairs of atoms and then tabulated for extrapolation. On the other hand, the key terms in the correction to the charge fluctuations in the Hamiltonian and in the total energy expression are approximated using exponentially decaying spherical charge densities. These two approximations used in the scheme are therefore independent and unrelated. Hence, although the scheme proposed in Ref. 11 is parameter-free, it may not be sufficiently flexible to yield a Hamiltonian with a wide range of transferability (see the discussion in Secs. II and IV).

**II. Methodology**

In this work, we present our scheme for the construction of semi-empirical and linear combination of atomic orbitals (LCAO)-based Hamiltonians for materials that allows the inclusion of a self-consistent (SC) determination of the charge redistribution and the environment-dependent (ED) multi-center interactions in a transparent manner. We chose a semi-empirical route for the construction of the system Hamiltonian because it has the flexibility to allow the database to provide the necessary ingredients for fitting parameters to capture the effect of electron screening. In the framework of a semi-empirical LCAO-based approach, the Hamiltonian is defined in terms of parameterized matrix elements $H_{i\alpha,j\beta}(\vec{R}_{ij})$ in some finite set of basis functions $\{\phi_{i\alpha}(\vec{r})\}$ not explicitly stated, where $i\alpha$ denotes the $\alpha$-orbital at the site $i$, and $\vec{R}_{ij} = \vec{R}_j - \vec{R}_i$ gives the relative position of the $j^{th}$ site with respect to the $i^{th}$ site. Within this context, the eigenvector $c_\lambda$, defining the coefficient vector of the expansion of the eigenfunction $\psi_\lambda$ in terms of $\{\phi_{i\alpha}(\vec{r})\}$, satisfies a general eigenvalue equation

$$Hc_\lambda = E_\lambda S c_\lambda \qquad (1)$$

with $S_{i\alpha,j\beta}(\vec{R}_{ij})$, the overlap matrix elements, being parameterized functions of $R_{ij}$ within the framework of the basis functions $\{\phi_{i\alpha}(\vec{r})\}$. Our strategy for developing a general scheme to construct a reliable and transferable SCED-LCAO Hamiltonian for materials with predictive power is given as follows:

The Hamiltonian of an aggregate of many-atom may be written as

$$H = -\sum_{l}\frac{\hbar^2}{2m}\nabla_l^2 + \sum_{l,i}v(\vec{r}_l - \vec{R}_i) + \sum_{l,l'}\frac{e^2}{4\pi\varepsilon_0 r_{ll'}} + \sum_{i,j}\frac{Z_i Z_j e^2}{4\pi\varepsilon_0 R_{ij}} \qquad (2)$$

where $r_{ll'} = |\vec{r}_l - \vec{r}_{l'}|$, $v(\vec{r}_l - \vec{R}_i)$ is the potential energy between an electron at $\vec{r}_l$ and the ion at $\vec{R}_i$, $Z_i$ the number of valence electrons associated with the ion at site $\vec{R}_i$, and the summation over $l$ and $l'$ runs over all the valence electrons. Within the one-particle approximation in the framework of linear combination of atomic orbitals, the on-site (diagonal) element of the Hamiltonian can be written as

$$H_{i\alpha,i\alpha} = \varepsilon_{i\alpha}^0 + u_{i\alpha}^{intra} + u_{i\alpha}^{inter} + v_{i\alpha} \qquad (3)$$

where $\varepsilon_{i\alpha}^0$ denotes the sum of the kinetic energy and the energy of interaction with its own ionic core of an electron in the orbital $i\alpha$. The terms $u_{i\alpha}^{intra}$ and $u_{i\alpha}^{inter}$ are the energies of interaction of the electron in orbital $i\alpha$ with other electrons associated with the same site $i$ and with other electrons in orbital $j\beta$ ($j \neq i$), respectively. The term $v_{i\alpha}$ represents the interaction energy between the electron in orbital $\alpha$ at site $i$ and the ions at the other sites. In our scheme, the terms in Eq. (3) are represented by

$$\varepsilon_{i\alpha}^0 = \varepsilon_{i\alpha} - Z_i U \qquad (4)$$

$$u_{i\alpha}^{intra} = N_i U \qquad (5)$$

and

$$u_{i\alpha}^{inter} + v_{i\alpha} = \sum_{k \neq i}[N_k V_N(R_{ik}) - Z_k V_Z(R_{ik})] \qquad (6)$$

where $\varepsilon_{i\alpha}$ may be construed as the energy of the orbital $\alpha$ for the isolated atom at $i$, $Z_i$ the number of positive charges carried by the ion at $i$ (also the number of valence electrons associated with the isolated atom at $i$), $N_i$ the number of valence electrons associated with the atom at $i$ when the atom is in the aggregate, $U$, a Hubbard-like term, the effective energy of electron-electron interaction for electrons associated with the atom at site $i$, $V_N(R_{ik})$ the effective energy of electron-electron interaction for electrons associated with different atoms (atoms $i$ and $k$), and $V_Z(R_{ik})$ the effective energy of interaction between an electron associated with an atom at $i$ and an ion at site $k$. In our approach, $\varepsilon_{i\alpha}$ may be chosen according to its estimated value based on the orbital $i\alpha$, or treated as a parameter of optimization. The quantity $U$ will be treated as a parameter of optimization while $V_N(R_{ik})$ and $V_Z(R_{ik})$ will be treated as parameterized functions to be optimized. An examination of Eqs. (3) through (6) clearly indicates that the presence of $N_i$, the charge distribution at site $i$, in the Hamiltonian provides the framework for a self-consistent determination of the charge distribution.

Following the same reasoning, we can set up the off-diagonal matrix element $H_{i\alpha,j\beta}$ ($j \neq i$) as

$$H_{i\alpha,j\beta} = \frac{1}{2}\{K(R_{ij})(\varepsilon'_{i\alpha} + \varepsilon'_{j\beta}) + [(N_i - Z_i) + (N_j - Z_j)]U$$

$$+ [\sum_{k \neq i}(N_k V_N(R_{ik}) - Z_k V_Z(R_{ik})) + \sum_{k \neq j}(N_k V_N(R_{jk}) - Z_k V_Z(R_{jk}))]\} S_{i\alpha,j\beta}(R_{ij}) \quad (7)$$

Thus, in addition to the conventional two-center hopping-like first term, Eq. (7) also includes both intra- and inter-electron-electron interaction terms as well as environment-dependent multi-center (three-center explicitly and four-center implicitly) interactions. From Eq. (7), it can be seen that the environment-dependent multi-center interactions are critically dependent on $V_N(R_{ik})$ and $V_Z(R_{ik})$, in particular their difference $\Delta V_N(R_{ik}) = V_N(R_{ik}) - V_Z(R_{ik})$. Since $V_Z(R_{ik})$ is defined as the energy of effective interaction per ionic charge between an ion at site $k$ and an electron associated with the atom at site $i$, we may model $V_Z(R_{ik})$ by the following parameterized function

$$V_Z(R_{ik}) = \frac{E_0}{R_{ik}}\{1 - (1 + B_Z R_{ik})e^{-\alpha_Z R_{ik}}\} \tag{8}$$

where

$$E_0 = \frac{e^2}{4\pi\varepsilon_0} \tag{9}$$

As both $V_N(R_{ik})$ and $V_Z(R_{ik})$ must approach $E_0/R_{ik}$ for $R_{ik}$ beyond a few nearest neighbor separations, $\Delta V_N(R_{ik})$ is expected to be a short ranged function of $R_{ik}$. We chose to model this short-ranged function by

$$\Delta V_N = (A_N + B_N R_{ik})\frac{[1 + e^{-\alpha_N d_N}]}{[1 + e^{-\alpha_N(d_N - R_{ik})}]} \tag{10}$$

on account of the flexibility of the expression given in Eq. (10). Since $V_N(R_{ik}) \to U$ as $R_{ik} \to 0$, Eqs. (8) and (10) then leads to

$$A_N = U - (\alpha_Z - B_Z)E_0 \tag{11}$$

In its broadest sense, the first term in Eq. (7) corresponds to the Wolfsberg-Helmholtz relation in the extended Hückel theory[13]. We modeled $K$ as a function of $R_{ij}$ rather than a constant parameter to ensure a reliable description of the dependence of the two-center term on $R_{ij}$ in the off-diagonal Hamiltonian matrix element. We found that an appropriate representation of $K(R_{ij})$ is given by

$$K(R_{ij}) = e^{\alpha_K R_{ij}} \tag{12}$$

The overlap matrix elements $S_{i\alpha, j\beta}(R_{ij})$ are expressed in terms of $S_{ij,\tau}$, with $\tau$ denoting,

for example, molecular orbitals $ss\sigma$, $sp\sigma$, $pp\sigma$, and $pp\pi$ in a $sp^3$ configuration. Since they are short-ranged function of $R_{ij}$, we chose to represent them by

$$S_{ij,\tau} = (A_\tau + B_\tau R_{ij})\frac{[1+e^{-\alpha_\tau d_\tau}]}{[1+e^{-\alpha_\tau(d_\tau - R_{ij})}]} \tag{13}$$

Based on the orthogonality of the $s$ and $p$ orbitals at the same site, we have

$$A_{ss\sigma} = A_{pp\sigma} = A_{pp\pi} = 1 \text{ and } A_{sp\sigma} = 0 \tag{14}$$

Equations (3) through (14) completely define the recipe for constructing semi-empirical SCED-LCAO Hamiltonians for materials in terms of parameters and parameterized functions. These parameters, including those characterizing the parameterized functions, are to be optimized with respect to a judiciously chosen database for a particular material.

The total energy of the system consistent with the Hamiltonian described by Eqs. (3) through (14) is given by

$$E_{tot} = E_{BS} - E_{corr} + E_{ion-ion} \tag{15}$$

where $E_{BS}$ is the band-structure energy and is obtained by solving the general eigenvalue equation (Eq. (1)), $E_{corr}$ is the correction to the double counting of the electron-electron interactions between the valence electrons in the band-structure energy calculation, and $E_{ion-ion}$ is the repulsive interaction between ions. Based on Eqs. (3) through (14), Eq. (15) can be rewritten as

$$E_{tot} = E_{BS} + \frac{1}{2}\sum_i (Z_i^2 - N_i^2)U - \frac{1}{2}\sum_{i,k(i\neq k)} N_i N_k V_N(R_{ik}) + \frac{1}{2}\sum_{i,k(i\neq k)} Z_i Z_k V_C \tag{16}$$

with

$$V_C = \frac{e^2}{4\pi\varepsilon_0 R_{ik}} = \frac{E_0}{R_{ik}} \tag{17}$$

It is illuminating to demonstrate how our approach relates to Frauenheim *et al*'s approach in Ref. 11. We may partition the SCED-LCAO Hamiltonian $H_{i\alpha,j\beta}$ into the two-center term $H^0_{i\alpha,j\beta}$ and the environment-dependent term such that

$$H_{i\alpha,i\alpha} = H^0_{i\alpha,i\alpha} + (N_i - Z_i)U + \sum_{k \neq i}[N_k V_N(R_{ik}) - Z_k V_Z(R_{ik})] \tag{18}$$

and

$$H_{i\alpha,j\beta} = H^0_{i\alpha,j\beta} + \frac{1}{2}\{[(N_i - Z_i) + (N_j - Z_j)]U$$

$$+ [\sum_{k \neq i}(N_k V_k(R_{ik}) - Z_k V_Z(R_{ik})) + (N_k V_N(R_{jk}) - Z_k V_Z(R_{jk}))]\}S_{i\alpha,j\beta}(R_{ij}) \tag{19}$$

where

$$H^0_{i\alpha,i\alpha} = \varepsilon_{i\alpha} \tag{20}$$

and

$$H^0_{i\alpha,j\beta} = \frac{1}{2}K(R_{ij})(\varepsilon'_{i\alpha} + \varepsilon'_{j\beta})S_{i\alpha,j\beta}(R_{ij}) \tag{21}$$

Since

$$E_{BS} = \sum_{i\alpha,j\beta}\sum_\lambda^{occ} c^\lambda_{i\alpha} c^\lambda_{j\beta} H_{j\beta,i\alpha} \tag{22}$$

the substitution of Eqs. (18) to (21) to Eq. (22) leads to

$$E_{BS} = E^0_{BS} + \sum_i (N_i - Z_i)U(\sum_{\alpha,\lambda}(c^\lambda_{i\alpha})^2 + \sum_{\alpha,j\beta,\lambda} c^\lambda_{i\alpha} c^\lambda_{j\beta} S_{j\beta,i\alpha})$$

$$+ \sum_{k \neq i}\sum_i [N_k V_N(R_{ik}) - Z_k V_Z(R_{ik})](\sum_{\alpha,\lambda}(c^\lambda_{i\alpha})^2 + \sum_{\alpha,j\beta,\lambda} c^\lambda_{i\alpha} c^\lambda_{j\beta} S_{j\beta,i\alpha}) \tag{23}$$

where

$$E_{BS}^0 = \sum_{i\alpha,j\beta} \sum_{\lambda}^{occ} c_{i\alpha}^{\lambda} c_{j\beta}^{\lambda} H_{i\alpha,j\beta}^0 \tag{24}$$

is the band structure energy corresponding to the two-center term. Recognizing that

$$N_i = \sum_{\alpha,\lambda} (c_{i\alpha}^{\lambda})^2 + \sum_{\alpha,j\beta,\lambda} c_{i\alpha}^{\lambda} c_{j\beta}^{\lambda} S_{j\beta,i\alpha}$$

we obtain

$$E_{BS} = E_{BS}^0 + \sum_i (N_i - Z_i) N_i U + \sum_{i,k(k \neq i)} [N_k V_N(R_{ik}) - Z_k V_Z(R_{ik})] N_i \tag{25}$$

The substitution of Eq. (25) into Eq. (16) yields

$$E_{tot} = E_{BS}^0 + \frac{1}{2} \sum_i (N_i - Z_i)^2 U + \frac{1}{2} \sum_{i,k(k \neq i)} N_i N_k V_N(R_{ik}) - \sum_{i,k(k \neq i)} N_i Z_k V_Z(R_{ik})$$

$$+ \frac{1}{2} \sum_{i,k(k \neq i)} Z_i Z_k V_c(R_{ik}) \tag{26}$$

Since

$$V_Z = V_N - \Delta V_N, \tag{27}$$

$$V_C = V_Z + \Delta V_C = V_N - \Delta V_N + \Delta V_C, \tag{28}$$

with

$$\Delta V_C = V_C (1 + B_Z R_{ik}) e^{-\alpha_Z R_{ik}}, \tag{29}$$

the substitution of Eqs. (27)-(28) into Eq. (26) yields

$$E_{tot} = E_{BS}^0 + \frac{1}{2} \sum_i \Delta N_i^2 U + \frac{1}{2} \sum_{i,k(k \neq i)} \Delta N_i \Delta N_k V_N(R_{ik})$$

$$+ \sum_{i,k(k \neq i)} N_i Z_k \Delta V_N + \frac{1}{2} \sum_{i,k(k \neq i)} Z_i Z_k (\Delta V_C - \Delta V_N) \tag{30}$$

Equation (30) indicates that the total energy in our approach can be expressed as the sum of the two-center band structure energy (first term), Coulomb-like energy associated with the charge fluctuations (second and third terms), and the short-ranged terms (fourth and fifth terms). It reduces to an expression similar to that of Ref. 11 only if we impose the condition $V_N = V_Z$ or $\Delta V_N = 0$, with $E_{rep} = \frac{1}{2}\sum_{i,k(k\neq i)} Z_i Z_k \Delta V_C$. Furthermore, Eq. (30) also shows that, even for systems with no charge redistribution, the total energy expression is different from that of Ref. 11 because of the presence of the term $\Delta V_N$ on account of $V_N \neq V_Z$ ($E_{SR}$=short-range energy $= \frac{1}{2}\sum_{i,k(k\neq i)} Z_i Z_k (\Delta V_C + \Delta V_N)$). In addition, for such systems, the SCED-LCAO Hamiltonian (see Eqs. (6) and (7)) still contains environment-dependent terms while the Hamiltonian of Ref.11 no longer has any. The presence of the environment-dependent terms in the Hamiltonian for systems with no on-site charge redistribution affects the distribution of the electrons among the orbitals even though the total charge associated with a given site is not changed. Therefore, the effect of the environment- dependency will be reflected in the band structure energy through the solution to the general eigenvalue equation (Eq. (1)) as well as the total energy. This is probably the reason why the results for high-coordinated crystalline phases based on the approach of Ref. 11 do not agree well with the DFT results (see the discussion in Sec. IV).

According to the strategy given above, the framework of the proposed semi-empirical SCED-LCAO Hamiltonian will allow the self-consistent determination of the electron distribution at site $i$. The inclusion of environment-dependent multi-center interactions (three-center explicitly and four-center interactions implicitly) will provide the proposed Hamiltonian with the flexibility of treating the screening effect associated with electrons which is important for the structure stability of narrow band solids such as d-band transition metals, while at the same time, handling the effect of charge redistribution for systems with reduced symmetry on equal footing. Furthermore, as described above, the Hamiltonian is set up in such a way that the physics underlying each term in the Hamiltonian is transparent. Therefore, it will be convenient to trace the underlying physics for properties of a system under consideration when such a Hamiltonian is used to investigate a many-atom aggregate and predict its properties. The salient feature of our strategy is that, with the incorporation of all the relevant terms discussed previously, there is no intrinsic bias towards ionic, covalent, or metallic bonding for the proposed Hamiltonian. Thus our strategy represents an approach that provides the appropriate conceptual framework to allow the chemical trend in a given atomic aggregate to determine the structural as well as electronic properties of condensed matter systems. In our strategy, there will be only about 20 fitting parameters in the construction of the proposed Hamiltonian for single component systems with a $sp^3$ basis. Our approach requires far less parameters, compared to phenomenological approaches[3,4] (with ~ 50 to 100 parameters) where environment-dependent effects are emphasized. In addition, the roles played by these parameters are well defined in terms of their physical significance. With far fewer parameters needed for the description of the proposed Hamiltonian, the optimization scheme for the determination of these parameters will be more robust. In our strategy, these parameters will be fitted to properties of stable configurations obtained from experiments and/or reliable first principles calculations, as well as metastable configurations determined by first principles calculations. Our approach differs from the DFT-based TB approach of Ref. 11 in the following important aspects. (1) A uniform treatment of the environment-dependent multi-center interactions for systems with or without the charge redistribution, resulting in a transferable Hamiltonian for a wide range of phases for materials beyond the scope of the approach in Ref. 11 as well as all other existing approaches. It should be noted that our treating environment-dependent interactions for systems with or without the charge redistribution on an equal footing highlights the important feature, the difference between $V_N(R_{ik})$ and $V_Z(R_{ik})$, that plays the crucial role in modeling the effects of electron screening in an atomic aggregate and that is completely ignored in the approach of Ref. 11. (2) A database-driven semi-empirical approach. Our approach depends critically on the database. If one can judiciously compile a systematic and reliable database, our scheme has the flexibility to allow the database to properly model the screening effect of the electrons in an atomic aggregate.

We have also implemented a MD scheme based on the SCED-LCAO Hamiltonian. In the MD simulations, the forces acting on the atoms in the atomic aggregate must be calculated at each MD step. The calculation of the band structure contribution to atomic forces can be carried out by Hellmann-Feynman theory[14]. With the presence of terms involving $N_i$ and $N_k$ in the SCED-LCAO Hamiltonian (see Eqs. (6) and (7)), terms such as $\nabla_k N_i$ where $\nabla_k$ refers to the gradient with respect to $\vec{R}_k$ will appear in the electronic contribution to the atomic forces. However, these terms are canceled exactly by terms arising from the gradients of the second and the third terms in the total energy expression (Eq. (16)). Thus terms involving $\nabla_k N_i$ will not contribute to the calculation of atomic forces. This fact greatly simplifies the calculation of atomic forces needed in the MD simulations. In other words, if one disregards the extra time due to the self-consistency requirement, the calculation of atomic forces based on the SCED-LCAO Hamiltonian is not anymore difficult compared with conventional TB approaches.

Finally, when there is charge redistribution, the Ewald method[15] can be used to calculate the long-range Coulomb interactions for extended systems. For finite systems, direct summation of the Coulomb terms can be used.

**III. Optimization of the parameters**

Our model consists of about 20 semi-empirical parameters $s_i$ for a single component system with a sp$^3$ basis for which the best or optimized values must be determined. The first step is to define a *residual* or *objective function R* which depends on the semi-empirical parameters, and for which the minimum value of $R$ is interpreted as the best value. We use a least-squares sum of the differences between the calculated properties $P_{calc}$ and the reference values $P_{ref}$:

$$R(s_i) = \sqrt{\frac{1}{N_P} \cdot \sum_k \left( P_{weight}^k \cdot \frac{P_{calc}^k(s_i) - P_{ref}^k}{P_{scale}^k} \right)^2} \qquad (31)$$

This expression also includes the characteristic scale $P_{scale}$ of each property, a weight factor $P_{weight}$ which represents the relative importance of each property, and the total number of properties $N_P$. The use of the scale, weight, and number of properties allows for the interpretation of the residual as the average relative deviation of the calculated values from the reference values.

The optimization problem is to find the *global* minimum, which is the set $s_i$ which has the absolute smallest value of $R$. Optimization algorithms however are fundamentally related to the number and distribution of *local* minima. In the easiest case, there would be only one local minimum, and only about $10^3$ evaluations of $R$ would be needed to find the global minimum. In the worst case, the local minima would be distributed randomly, and only a brute-force search could find the global minimum. In this worst case scenario, the number of function evaluations needed would be $G_s^{N_s}$, where $G_s$ is the number of points for each parameter, and $N_s$ is the number of parameters. A reasonable value of $G_s$ is ~200, and with ~20 semi-empirical parameters it is evident that the optimization problem would be intractable. Unfortunately, it seems that the atomic-scale modeling problem, while not intractable, is still of the more difficult type. This means that the selection of the optimization algorithm, and also the selection of the initial or starting values of the semi-empirical parameters, is particularly important.

For the initial parameters, we use results adapted from the available literature. First-principles calculations of the two-center interactions such as $S_{ss\sigma}$ are available for $Si$[16]. For the least squares problem with $N_P$ on the order of $10^2$, there are on the order of $10^2$ terms in the summation. If this summation is performed explicitly, a large amount of information about the individual behavior of these terms is lost. For example, if the summation is performed explicitly, the only information available about the derivatives is the gradient $\frac{\partial R}{\partial s_i}$ with $N_s$ elements, but if the summation is not performed explicitly then the entire Jacobian $\frac{\partial P_k}{\partial s_i}$ with $N_s \cdot N_p$ elements is available. So even though the problem is to find the minimum value of $R$, efficient least-squares algorithms do not perform the summation explicitly, but rather store and analyze each of the $10^2$ terms in the summation. This is the approach used by the Marquardt-Levenberg algorithm[17], which is a widely used and highly efficient algorithm for finding the local minimum of a least-squares problem. When compared with any algorithm which analyzes only the value of $R$, least-squares algorithms are typically one or two orders of magnitude more efficient at finding the local minimum, with the efficiency increasing for larger values of $N_P$.

Now, the least-squares problem and the Marquardt-Levenberg algorithm are well-understood[17]. Also, the global optimization problem for a scalar function is well-understood[17]. However, we have here a *global least-squares* problem. This problem is not well-understood and is rarely discussed in the literature. There are two general approaches to the global least-squares problem. The first is to treat the least-squares problem as a scalar optimization problem, analyzing only the value of $R$ and not the values of the individual least-squares terms. The reasoning here is that the benefit of using pre-existing and well-understood algorithms, such as a simulated annealing algorithms, will outweigh the cost of not analyzing the individual terms in the summation. The second approach is to adapt a local least-squares algorithm to the global problem. Here one can exploit the superior efficiency of the least-squares algorithm, although there is now the cost of developing on your own the global part of the algorithm.

Our experience indicates that the second approach of treating the problem as a least-squares problem is considerably more efficient. This is probably related to the fact that we are now using up to 200 properties for the least-squares summation, which is considerably more than have been used previously. We have developed our own global adaptation of the Marquardt-Levenberg algorithm, which is quite efficient for the global least-squares problem. The details will be published elsewhere. Specifically, we feed successive sets of parameters $s_i$ to the Marquardt-Levenberg algorithm, which finds the local minimum for each set of parameters. Each successive set is chosen with a random distribution from the best set found from all the previous local optimizations. For the random distribution we define a scalar "distance" $s$ from the next set of parameters $s_{next}$ to the best set of parameters $s_{best}$ as:

$$s = \sqrt{\frac{1}{N_s} \cdot \sum_i \left( \frac{s_{next}^i - s_{best}^i}{s_{scale}^i} \right)^2} \qquad (32)$$

where $s_{scale}$ is the characteristic scale of each parameter, and $N_s$ is the total number of parameters. We then assign a value for $s$ using a random exponential distribution. From the best parameters $s_{best}$ and the random distance $s$ it is then a straightforward matter to construct the next set of parameters $s_{next}$ to feed to the local least-squares algorithm. The explicit algorithm for $s_{next}^i$ is then:

$$s_{next}^i = s_{best}^i + \frac{-s_{global} \cdot \ln(r_{0,1})}{\sqrt{\frac{\sum_k r_{-1,+1}^k \cdot r_{-1,+1}^k}{N_s}}} \cdot r_{-1,+1}^i \cdot s_{scale}^i \qquad (33)$$

EQ $r_{a\backslash b}$ is a random number with a uniform distribution over the interval $[a,b]$, and $s_{global}$ a unitless number which represents the expected range over which the local minima are distributed. A typical value of $s_{global}$ is 0.5, which means that the new parameters will differ from the old parameters by about 50%. Although it is evident from the context of the discussion, we should emphasize that the array $r^i$ or $r^k$ refers to the same array of random numbers for each of its 3 appearances in this equation. Loosely speaking, the random exponential distribution means that the new set of parameters is more likely to be close to the best set of parameters.

Perhaps the most important feature of this random distance algorithm is that successive sets are chosen with regard to the values of the parameters $s$ and without regard to the residual value $R$. This is in contrast with techniques which interpret the residual $R$ as a type of energy barrier. The highly nonlinear behavior of the residual even in regions where the parameters are reasonable suggests that simulated-annealing-type techniques are not appropriate for these types of optimization problems because the residual barriers are too large. If one does wish to adapt this algorithm to problems where a simulated-annealing-type interpretation is more appropriate, $s_{best}$ can be allowed to "hop" to a local minimum which is not necessarily the best local minimum, with a hopping probability that depends on the difference between two appropriate $R$ values. Indeed, we have used this simulated-annealing-type adaptation at times, and although it certainly adds flair to the algorithm, it does not seem to be useful for our particular problem.

**IV. Case studies: application to silicon-based structures**

We have tested our strategy of constructing the semi-empirical SCED-LCAO Hamiltonian for materials, using silicon as our working example because of the existence of the large body of experimental as well as theoretical information on silicon-based structures. The parameters characterizing the SCED-LCAO Hamiltonian for silicon (see Sec. II), obtained using the optimization procedure outlined in Sec. III, are given in Table 1. The properties used to determine this set of parameters include: The binding energies and bond lengths for $Si_n$ clusters with n=2 to 6 (see Table 2)[18]; the energy vs. atomic volume curves for the diamond, the simple cubic (sc), the body centered cubic (bcc), and the face centered cubic (fcc) phases[19]; the band structure energies at high symmetry points for the diamond phase[19-22] (see Table 3).

The results showing the energy vs atomic volume curves for the diamond, the simple cubic (sc), the body centered cubic (bcc), and the face centered cubic (fcc) phases of silicon, obtained by using the SCED-LCAO Hamiltonian constructed for Si with our scheme, are presented in Fig. 1. Also shown in Fig. 1 are the corresponding curves obtained using three existing traditional (two-center and non-self consistent) non-orthogonal tight binding (NOTB) Hamiltonians[16,23,24], and two more recently developed non-self consistent but environment-dependent Hamiltonians[7,9]. All the curves (solid) are compared with the results obtained by DFT-LDA calculations[19] (dotted). It can be seen that while the results obtained by all the existing Hamiltonians fail for the high pressure phases, those obtained using Hamiltonians with environment-dependent terms give much better agreement for those phases. This is an indication of the importance of the inclusion of the environment-dependent effects in the Hamiltonian, even for single-element extended crystalline phases. However, the most striking message conveyed by Fig.1 is how well our result compares with the DFT-LDA results for all the extended crystalline phases, both at low as well as high pressures. It indicates that our scheme has the capacity and the flexibility of capturing the environment-dependent screening effect under various local configurations. We have also checked the self-consistency in the charge redistribution by using the SCED-LCAO Hamiltonian to study the structural properties of $Si_n$ clusters with n ranging from 2 to 6. The results on the binding energy and bond lengths for the stable and meta-stable structures of these clusters all agree excellently with the first principles results[18] (see Table 2).

To test the robustness of the self-consistent scheme and demonstrate the predictive power of the SCED-LCAO Hamiltonian, we have applied the SCED-LCAO Hamiltonian to study three diverse examples of silicon-based structures of reduced or no symmetry. The results are given as follows.

## A. Structural Properties of Si$_{71}$ Cluster

We have used the MD scheme based on the SCED-LCAO Hamiltonian to determine the stable structure of Si$_{71}$, an intermediate-size cluster. We generated the initial configuration of Si$_{71}$ cluster from the truncated tetrahedral network. We first heated and equilibrated this initial configuration at 500 K for about 2.4 ps. We then annealed it to 300 K for about 0.7 ps, and finally cooled it down to 0 K for about 2 ps.

The atoms on the truncated "surface" of the initial tetrahedral configuration of the Si$_{71}$ cluster have many dangling bonds. Therefore the initial configuration is very unstable. Two factors that play key roles in determining a stable configuration of a Si cluster are: (1) saturation of the dangling bonds of the surface atoms; (2) the tendency to maintain the coordination number for Si atoms closer to four. The interplay of these factors will lead to a distorted surface for a stable Si cluster. This can be seen from Fig. 2b where the stabilized Si$_{71}$ cluster obtained by our simulation shows a compact network in a more oblate structure. The strong surface distortion is a reflection of local bonding configurations with the number of bonds associated with atoms in the cluster, in particular the surface atoms, close to four. This type of structures has also been found to be more stable for other Si clusters of intermediate size by previous theoretical studies[25-28].

We have also calculated the pair distribution function, $g(r)$, for the equilibrated Si$_{71}$ cluster shown in the inset of Fig. 2. From Fig. 2, it can be seen that $g(r)$ exhibits a very sharp first peak followed by a broader second peak, a typical feature of distorted cluster structure. Also shown in Fig. 2 is the pair distribution function for the stable Si$_{71}$ cluster obtained under the same equilibration procedure but using the DFT-based Fire-Ball MD scheme[29]. It can be seen that the agreement between the result from the SCED-LCAO MD scheme and that from the Fire-Ball MD scheme is excellent.

It is well known that the charge redistribution plays the critical role in establishing chemical bonding in relaxation. This is particularly true for "surface" atoms in a cluster of intermediate size. The result of our test case therefore has demonstrated the robustness of the self-consistent scheme in the determination of the charge redistribution in the SCED-LCAO Hamiltonian.

**B. Reconstruction of the Si(001) Surface**

We have carried out a MD simulation of the reconstruction of Si(001) surface from scratch, using the SCED-LCAO Hamiltonian. We started with the ideal Si(001) with P1x1 symmetry as the initial configuration. We chose a 4x4 slab with a thickness of 8 layers as the MD cell. In the simulation, the atoms in the top 4 layers were allowed to fully relax while the atoms in the bottom 4 layers were kept at their bulk equilibrium positions. We turned on the simulations by first moving the surface atoms in the alternate column towards the fixed surface atoms by $\leq 0.1$ Å.

We found that the surface reconstruction of the Si(001) surface takes about 0.5 ps (see Fig. 3). The surface atoms begin to dimerize in ~0.05 ps after performing the SCED-LCAO MD relaxation. These dimers become buckled (tilted) after about another 0.075 ps. Finally the surface reconstruction stabilizes to the stable configuration with the C4x2 symmetry in another ~0.375 ps. To the best of our knowledge, this is the first time that the C4x2 reconstruction of the Si(001) surface is obtained directly from the dynamical relaxation simulation of its ideal P1x1 surface configuration. It demonstrates the predictive power of the SCED-LCAO Hamiltonian. It should be noted that the C4x2 configuration can not be obtained when the simulation is performed without the self-consistent requirement of the charge, indicating that charge redistribution is a key ingredient during the surface reconstruction.

In Table 4, the properties characterizing the buckled C4x2 reconstruction of the Si(001) surface obtained by the SCED-LCAO Hamiltonian-based MD simulation are compared with the corresponding properties obtained by DFT calculation[30] and/or experimental measurements[31]. It can be seen that the agreement is very good.

**C. Mapping the Energy Landscape of a Si-adatom on the reconstructed Si(111)-(7x7) Surface**

Finally, we have applied the SCED-LCAO Hamiltonian to map out the energy landscape for a Si adatom on the reconstructed Si(111) surface. This represents a most stringent test for the reliability and efficiency of the application of the SCED-LCAO Hamiltonian because of the complicated reconstruction of the Si(111) surface. In our study, we used the SCED-LCAO-MD scheme to unravel the structural and the dynamical behavior of an adsorbed Si atom on the Si(111)-(7x7) dimer-adatom-stacking-fault (DAS)-reconstructed surface[32]. To have a complete understanding of the behavior pattern of the Si adatom on the Si(111)-7x7 surface, we included both the faulted and the unfaulted halves of the Si(111)-7x7 DAS structure. Therefore, by necessity, we have used a large supercell composed of 10 layers plus the adatom layer (494 atoms in total), where the top 8 layers were relaxed and the bottom two layers were held at their bulk equilibrium positions.

The preferential adsorption sites for the adsorbed Si atom were established by mapping out the total energy as a function of its positions. In Figs. 4a and 4b, the adsorption energy along two pathways in the faulted half of the unit cell is shown respectively. These two pathways are composed of irreducible sites in the faulted half. As shown in Table 5, the calculated adsorption energies for sites along the two pathways in the faulted half exhibit many stable adsorption sites (T4-CE, T2-CE, B2-CEA1, B2-CEA2, B2-COA, and H3-COA along the path 1 and T2-CEA, T4-DR, T2-COA1, T2-COA2, O, and CH along the path 2). It is interesting to note that the stable adsorbate site is not on top of the rest atom (T1) or on top of the dimers (D2). The factor determining the stable adsorbate sites depends on the situation when, in addition to saturating any dangling bond of the surface atoms, the Si adsorbate atom can form more bonds with the substrate atoms so that its coordination number is closer to four. The adatom at site T1, although it saturates one dangling bond of the rest atom, does not satisfy the optimally coordinated criterion for silicon. Our calculation also reveals several low-energy barriers in both pathways, in particular, energy barriers of $\leq 0.3$ eV between the sites T2-CE and B2-CEA1, B2-CEA1 and B2-CEA2, or B2-CEA2 and B2-COA, or two equivalent B2-COAs in pathway 1 and between the sites T2-COA1 and T2-COA2, or between two equivalent T2-CEA sites in the pathway 2. These results are consistent with the result of theoretical calculations using the DFT-based VASP package[33].

Based on the energy landscape, the low barrier energies, and the fact that the sites are located close to each other, one can expect the silicon adatom to be trapped in one of the three types of basins of attraction in the faulted half described as follows (see Fig. 5).
**(1) Triangular-Type Basin**: The energy landscape calculation reveals three triangular-type basins of attraction surrounding the T1 sites on top of the rest atom, formed by sites of B2-, H3-, and T4-type as shown in Fig. 5. In each of the basin of attraction, the adsorption energy near the corner adatoms (i.e. the B2-COA and H3-COA sites) is lower than that near the central adatoms (i.e. the B2-CEA and H3-CEA sites). This anisotropy in energy in the triangular type of basin is consistent with the atom tracking image of an adsorbed Si atom at low temperatures (see Fig. 4(a) of Ref. 34), where it is reported that the adsorbed atom spends most of the time in the region defined by the positions R1, R2, R3 which are near the rest atoms and the corner adatoms COA1, COA2 and COA3.

**(2) Hexagonal Ring-Type Basin**: As shown in Fig. 5, a hexagonal ring-type basin of attraction is located at the center of the half unit cell and is composed of the T2-CE and T4-CE sites surrounding the H3-CE site with the T4-CE site having the lowest energy. This type of the basin of attraction provides the explanation for the atom-tracking image of an adsorbed Si atom at room temperature[34], where it is reported that the adsorbed atom spends most of the time inside the central region defined by the three center adatoms (CEA1, CEA2, and CEA3), but occasionally moves near the rest atom positions (R1, R2, or R3) and corner adatom (COA1 or COA2) positions, as shown in Fig. 2(a) of Ref. 34.

**(3) Shoulder-Type Basin**: The energy landscape calculated along pathway-2 reveals shoulder-type basin of attraction in the vicinity of the dimer row formed by the O, T2-CEA, T4-DR, T2-COA1, T2-COA2, and CH sites, as shown in Fig. 5. The T2-COA sites near the corner holes are lower in energy. This may explain the formation of the Si tetramers located on the top of the corner dimer at low temperature and on the top of central dimer at room temperature[34,35].

The combination of the three types of basins of attraction results in an attractive potential well that traps the adsorbed Si atoms to form magic clusters. In particular, the region bounded by T2-CE, T2-CE, T1, T2-CEA, T2-CEA, and T1 matches very well the schematic drawing of the six protrusions depicting the magic cluster on the faulted half of Si(111)-7x7 surface as noted by Hwang *et al*.[36] (see Fig. 5, the region bonded by T2-CE, T2-CE, T1, T2-CEA, T2-CEA, and T1). Furthermore, the low energy barriers allows the cluster to move within the half unit cell.

We have also compared adsorption energies of corresponding sites in the faulted half and the unfaulted half of the unit cell. We found that the adsorption energy of most of sites in the faulted half is lower compared with the corresponding site in the unfaulted half (see Table 5). Specifically, the sites with the two lowest energy, B2-COA and T4-CE, have lower energy in the faulted half than in the unfaulted half. This appears to be one of the reasons why the Si magic cluster prefers to form on the faulted half of the unit cell of Si (111)-(7x7) surface as observed by Hwang *et al*[36].

The application of the SCED-LCAO Hamiltonian for silicon to the three test cases discussed above represents a concerted effort to test the versatility, the reliability, and the efficiency of using the SCED-LCAO Hamiltonian to study properties of complex systems with no or reduced symmetry. The first case concerns a finite system with no symmetry. The second and third cases deal with extended systems with reduced symmetry. The properties of all these three low-dimensional systems are, therefore, critically dependent on charge re-distribution and environment-dependent multi-center interactions. The result of our test studies has clearly demonstrated that (i) the SCED-LCAO Hamiltonian for silicon is transferable and hence it has the predictive power; (ii) the self-consistent scheme for the determination of the charge re-distribution is robust; (iii) the MD code based on the SCED-LCAO Hamiltonian is efficient.

**V. Discussion**

The development of a recipe to construct semi-empirical Hamiltonians for elemental materials in the present work is grounded in the ingredients of the many-body Hamiltonians describing the many-atom aggregates. The determination of the parameters characterizing the SCED-LCAO Hamiltonians is database driven. In this sense, the SCED-LCAO Hamiltonian is only as good as the database used to optimize the fitting parameters. Our case studies on silicon-based structures indicates that, with the compilation of a judiciously chosen database, the resulting SCED-LCAO Hamiltonian is versatile, reliable, efficient and possesses predictive power. Thus, we are confident that reliable and transferable SCED-LCAO Hamiltonians with predictive power can be developed for real materials using our scheme. Furthermore, our scheme is efficient so that simulations of complex systems with large degrees of freedom can be conveniently carried out.

Construction of SCED-LCAO Hamiltonians for other column IV elements (e.g. carbon and germanium), simple metals (e.g., aluminum), and transition metals (e.g., iron and nickel) are currently in progress. The extension of the present scheme to its spin-polarized version is also in progress.

**Acknowledgement**

This work was supported by NSF (DMR 0112824) and DOE (DE-FG02-00ER45832). We would also like to acknowledge a helpful discussion with Prof. Weitao Yang.

# References

1. J.A. Majewski and P. Vogl, Phys. Rev. B**35**, 9666 (1987).

2. R.E. Cohen, M.J. Mehl, and D.A. Papaconstantopoulous, Phys. Rev. B**50**, 14694 (1994).

3. M.S. Tang, C.Z. Wang, C.T. Chen, and K.M. Ho, Phys. Rev. B**53**, 979 (1996).

4. A.F. Kohan and G. Ceder, Phys. Rev. B**54**, 805 (1996).

5. Q. Xie and P. Chen, Phys. Rev. B**56**, 5235 (1997).

6. C.M. Goringe, D. R. Bowler, and E. Hernandez, Rep. Prog.. Phys. **60**, 1447 (1997).

7. D.A. Papaconstantopoulos, M.J. Mehl, S.C. Erwin, and M.R. Pederson, in *Tight-Binding Approach to Computational Materials Science*, edited by P.E.A. Turchi, A. Gonis, and L. Colombo, MRS Symposia Proceedings No. 491 (Materials Research Society, Pittsburg, 1998), p. 221.

8. K. Esfarjani and Y. Kawazoe, J. Phys.: Condens. Matter **10**, 8257 (1998).

9. C.Z. Wang, B.C. Pan, and K.M. Ho, J. Phys.: Condens. Matter **11**, 2043 (1999).

10. A.N. Andriotis and M. Menon, Phys. Rev. B**59**, 15942 (1999).

11. Th. Frauenheim *et al*., phys. Stat. Sol. (b)**217**, 41 (2000).

12. W.M.C. Foulkes and R. Haydock, Phys. Rev. B**39**, 12520 (1989).

13. R. Hoffmann, J. Chem. Phys. **39**, 1397 (1963).

14. R.P. Feynman, Phys. Rev. **56**, 340 (1939).

15. J. Ihm, A. Zunger, and M.L. Cohen, J. Phys. C**12**, 4409 (1979); M.T. Yin and M.L. Cohen, Phys. Rev. B**26**, 3259 (1982).

16. Th. Frauenheim *et al*., Phys. Rev. B**52**, 11492 (1995).

17. J.E. Dennis Jr and R.B. Schnabel, *Numerical Methods for Unconstrained optimization and Nonlinear Equation*, (Society fro Industrial & Applied Mathmatics, 1996).

18. Gaussian 98 MPW1PW91/cc-pVTZ.

19. M.T. Yin and M.L. Cohen, Phys. Rev. B**26**, 5668 (1982).

20. W.D. Grobman and D.E. Eastman, Phys. Rev. Lett. **29**, 1508 (1972); W.D. Grobman, D.E. Eastman, and J.E. Freeouf, Phys. Rev. B**12**, 405 (1975).

21. L. Ley, S. Kowalczyk, R. Pollack, and D.A. Shirley, Phys. Rev. Lett. **29**, 1088 (1972).

22. W.E. Spicer and R. Eden, in *Proceedings of the Ninth International Conference of the Physics of Semiconductors*, (Navka, Leningrad, 1968), Vol. 1, P. 61.

23. M. Menon and K.R. Subbaswamy, Phys. Rev. B**55**, 9231 (1997).

24. N. Bernstein and E. Kaxiras, Phys. Rev. B**56**, 10488 (1997).

25. K.M. Ho *et al*., Nature **392**, 582 (1998).

26. J. Song, S.E. Ulloa, and D. Drabold, Phys. Rev. B**53**, 8042 (1996).

27. E. Kaxiras and K. Jackson, Phys. Rev. Lett. **71**, 727 (1993).

28. L. Mitas, J.C. Grossman, I. Stich, and J. Tobik, Phys. Rev. Lett. **84**, 1479 (2000).

29. O.F. Sankey and D.J. Niklewski, Phys. Rev. B**40**, 3979 (1989); A.A. Demkov, J. Ortega, O.F. Sankey, and M.P. Grumbach, Phys. Rev. B**52**, 1618 (1995).

30. J.E. Northrup, Phys. Rev. B**47**, 10032 (1993).

31. C.M. Wei, H. Huang, S.Y. Tong, G.D.S. Glander, and M.B. Webb, Phys. Rev. B**42**, 11284 (1990).

32. K. Takayanag *et al*., Surf. Sci. **164**, 367 (1985); K. Takayanag *et al*., J. Vac. Sci. Technol. A**3**, 1502 (1985).

33. C.M. Chang and C. M. Wei, Phys. Rev. B**67**, 033309 (2003).

34. T. Sato, S. Kitamura, and M. Iwatsuki, J. Vac. Sci. Technol. A**18**, 960 (2000).

35. T. Sato, S. Kitamura, and M. Iwatsuki, Surf. Sci. **445**, 130 (2000).

36. I.-S. Hwang, M.-S. Ho, and T.T. Tsong, Phys. Rev. Lett. **83**, 120 (1999).

Table 1: The parameters characterizing the SCED-LCAO Hamiltonian for silicon.

| Symbols | Values | Symbols | Values | Symbols | Values |
| --- | --- | --- | --- | --- | --- |
| $U$ | 8.05 eV | $\alpha_N$ | 2.74 Å$^{-1}$ | $\alpha_{sp\sigma}$ | 2.18 Å$^{-1}$ |
| $\varepsilon'_s$ | -13.43 eV | $d_N$ | 1.91 Å | $\alpha_{pp\sigma}$ | 2.35 Å$^{-1}$ |
| $\varepsilon'_p$ | -7.91 eV | $B_{ss\sigma}$ | 0.88 Å$^{-1}$ | $\alpha_{pp\pi}$ | 3.74 Å$^{-1}$ |
| $\alpha_K$ | 0.25 Å$^{-1}$ | $B_{sp\sigma}$ | -0.75 Å$^{-1}$ | $d_{ss\sigma}$ | 1.32 Å |
| $B_Z$ | 1.54 Å$^{-1}$ | $B_{pp\sigma}$ | -0.79 Å$^{-1}$ | $d_{sp\sigma}$ | 1.35 Å |
| $A_N$ | -1.26 eV | $B_{pp\pi}$ | -0.31 Å$^{-1}$ | $d_{pp\sigma}$ | 2.03 Å |
| $B_N$ | 0.16 eV Å$^{-1}$ | $\alpha_{ss\sigma}$ | 3.04 Å$^{-1}$ | $d_{pp\pi}$ | 2.28 Å |

Table 2: Comparison of the geometries (positive numbers in Å) and the binding energies (negative numbers in eV) of the Si clusters obtained in the present work with those by an *ab initio* calculation.

| Cluster | Symmetry | Present work | ab initio values[a] |
|---------|----------|--------------|---------------------|
| $Si_2$  | $D_{ih}$ | 2.226        | 2.288               |
|         |          | -2.435       | -2.499              |
| $Si_3$  | $C_{2v}$ | 2.284        | 2.357               |
|         |          | 2.168        | 2.158               |
|         |          | -3.413       | -3.575              |
| $Si_3$  | $D_{ih}$ | 2.141        | 2.167               |
|         |          | -3.427       | -3.404              |
| $Si_4$  | $D_{2h}$ | 2.275        | 2.311               |
|         |          | -4.101       | -4.242              |
| $Si_4$  | $T_d$    | 2.332        | 2.474               |
|         |          | -3.773       | -3.659              |
| $Si_4$  | $D_{ih}$ | 2.116        | 2.156               |
|         |          | 2.164        | 2.176               |
|         |          | -3.289       | -3.364              |
| $Si_5$  | $D_{3h}$ | 2.207        | 2.306               |
|         |          | 3.141        | 3.064               |
|         |          | -3.352       | -3.340              |
| $Si_5$  | $C_{4v}$ | 2.209        | 2.275               |
|         |          | 2.358        | 2.513               |
|         |          | -4.327       | -4.266              |
| $Si_5$  | $D_{ih}$ | 2.082        | 2.133               |
|         |          | 2.128        | 2.144               |
|         |          | -3.545       | -3.534              |
| $Si_5$  | $T_d$    | 2.127        | 2.215               |
|         |          | 3.475        | 3.617               |
|         |          | -3.334       | -3.383              |
| $Si_6$  | $D_{4h}$ | 2.248        | 2.363               |
|         |          | 2.639        | 2.734               |
|         |          | -4.698       | -4.664              |
| $Si_6$  | $D_{3d}$ | 2.261        | 2.285               |
|         |          | 2.948        | 3.208               |
|         |          | -3.896       | -3.972              |
| $Si_6$  | $D_{ih}$ | 2.057        | 2.098               |
|         |          | 2.072        | 2.134               |
|         |          | 2.149        | 2.158               |
|         |          | -3.446       | -3.446              |

a: reference 18

Table 3: Comparison of the band structure energies of bulk Si at high symmetry points (in eV) obtained in the present work (optimized lattice constant is 5.4464 Å) with those by a DFT calculation and from the angle-integrated photoemission spectra.

| Band index | Present work | DFT calculation | Experiment |
| --- | --- | --- | --- |
| $\Gamma_{1v}$ | -11.77 | -11.93[a] | -12.4 ± 0.6[b]; -12.5 ± 0.6[c] |
| $X_{4v}$ | -3.30 | -2.88[a] | -2.5 ± 0.3[c]; -2.9[d] |
| $L_{2'v}$ | -10.10 | -9.52[a] | -9.3 ± 0.4[c] |
| $L_{1v}$ | -6.62 | -7.00[a] | -6.4 ± 0.4[b]; -6.8 ± 0.2[c] |
| $L_{3v}$ | -1.89 | -1.20[a] | -1.2 ± 0.2[d] |

a: reference 19
b: reference 20
c: reference 21
d: reference 22

Table 4: Properties of buckled dimer row on the Si (001) C4x2 reconstructed surface, where $\Delta E$/dimer is the binding energy per dimer (in eV), b denotes the dimer bond length (in Å), $\Delta z$ the height of the bulked dimer (in Å), and α the angle of the dimer with respect to the surface (in degree).

| Properties | Present work | DFT-LDA | Experiment |
| --- | --- | --- | --- |
| $\Delta E$/dimer | 1.18 | 1.39[a] | |
| b | 2.47 | 2.29[a] | 2.45 ± 0.1[b] |
| $\Delta z$ | 0.69 | 0.69[a] | |
| α | 16.19 | 17.5[a] | |

a: reference 30
b: reference 31

Table 5: Adsorption energy in the irreducible region of Si(111)-(7x7) reconstructed surface. The site symbols are described as follows: T1 denotes an adsorption site on top of the rest atom with one dangling bond, T2 a site on top of layer-1 atom which is different from the rest atom, T4 a four-fold site on top of an un-dimerized atom of layer-2, H3 a hexagonal three-fold site, B2 a two-fold site between T4 and H3 or T2 and T4, D2 a site on top of a dimmer atom, P a site within the pentagonal ring and whose image site in the unfaulted half lies above a layer-4 atom, O a site within an octagonal ring and whose image site in the unfaulted half lies above a layer-4 atom, CH a site within the corner hole region and whose image site in the unfaulted half lies above a layer-4 atom, CEA a site on top of a central adatom, COA a site on top of a corner adatom, and COH the central position of the corner hole, respectively. In addition, the auxiliary notation CE denotes a site located in the central region of the half unit cell, CEA a site located near the central adatom, COA a site located near the corner adatom, and DR a site located near the dimer row, respectively.

| Site number along path 1 | $E_{adsorption}$ (eV) (faulted) | $E_{adsorption}$ (eV) (unfaulted) | Symbol of site type |
|---|---|---|---|
| 1 | -3.12048 | -3.29868 | T1 |
| 2 | -4.15503 | -4.00158 | T4-CE |
| 3 | -3.55113 | -3.35263 | H3-CE |
| 4 | -3.91248 | -3.90753 | T2-CE |
| 5 | -3.64064 | -3.47482 | B2-CE |
| 6 | -3.97138 | -4.06098 | B2-CEA1 |
| 7 | -3.70458 | -3.49668 | H3-CEA |
| 8 | -3.95703 | -3.97188 | B2-CEA2 |
| 9 | -3.77883 | -3.88883 | T4-DR |
| 10 | -4.22929 | -4.11543 | B2-COA |
| 11 | -3.94705 | -3.37095 | H3-COA |
| Site number along path 2 | $E_{adsorption}$ (eV) (faulted) | $E_{adsorption}$ (eV) (unfaulted) | Symbol of site type |
| 1 | -3.04120 |  | D2-CEA |
| 2 | -3.52638 | -3.37293 | P-CEA |
| 3 | -2.67844 | -2.68537 | CEA |
| 4 | -3.76398 | -3.78279 | T2-CEA |
| 5 | -3.77883 | -3.77883 | T4-DR |
| 6 | -4.08573 | -4.03078 | T2-COA1 |
| 7 | -2.97190 | -3.01653 | COA |
| 8 | -3.88773 | -3.81348 | T2-COA2 |
| 9 | -3.85308 | -3.44718 | CH |
| 10 | -2.93238 |  | COH |
| 11 | -3.33332 |  | D2-COA1 |
| 12 | -3.53133 | -3.75408 | P-COA |
| 13 | -3.20463 |  | D2-COA2 |
| 14 | -3.83320 | -3.14028 | O |

**Figure captions**

Figure 1  The energy vs atomic volume curves for the diamond (cdia), the simple cubic (sc), the body centerd cubic (bcc), and the face centered cubic (fcc) phases of silicon, obtained using the present SCED-LCAO scheme (top-left plot).  The corresponding curves obtained using three existing traditional (two-center and non-self consistent) non-orthogonal tight binding (NOTB) Hamiltonians (top-central[23], top-right[24], and bottom-left[16] plots), and two more recently developed non-self consistent but environment-dependent Hamiltonians (bottom-central[9] and bottom-right[7] plots) are also shown in the figure. All the curves (solid) are compared with the result obtained by a DFT-LDA calculation[19] (dotted).

Figure 2  The pair-distribution function g(r) for the equilibrated $Si_{71}$ cluster (with its structure shown in the inset) obtained by SCED-LCAO (solid) is compared with that obtained by a DFT-LDA   calculation[29] (dotted).

Figure 3  The Si (001) surface reconstruction simulated starting from the ideal P1x1 symmetry (left inset) and stabilizing to the C4x2 symmetry (right inset) is shown in term of the total energy per atom vs the molecular dynamical time step.

Figure 4  The irreducible sites (stars) along the first pathway (a) and the second pathway (b) in the faulted half. The symbols corresponding to site numbers are also shown. The triangular region bonded by the dashed line is the irreducible region of the Si(111)-(7x7) reconstructed surface.

Figure 5  The sites having lower adsorption energy on the Si(111)-(7x7) reconstructed surface. The solid, dotted, and dotted-dash curves are guides to the triangular-type, hexagonal ring-type, and shoulder-type of basins of attraction, respectively. The T1 and T2-CEA define the region of experimentally observed six protrusions associated with the magic clusters[36].

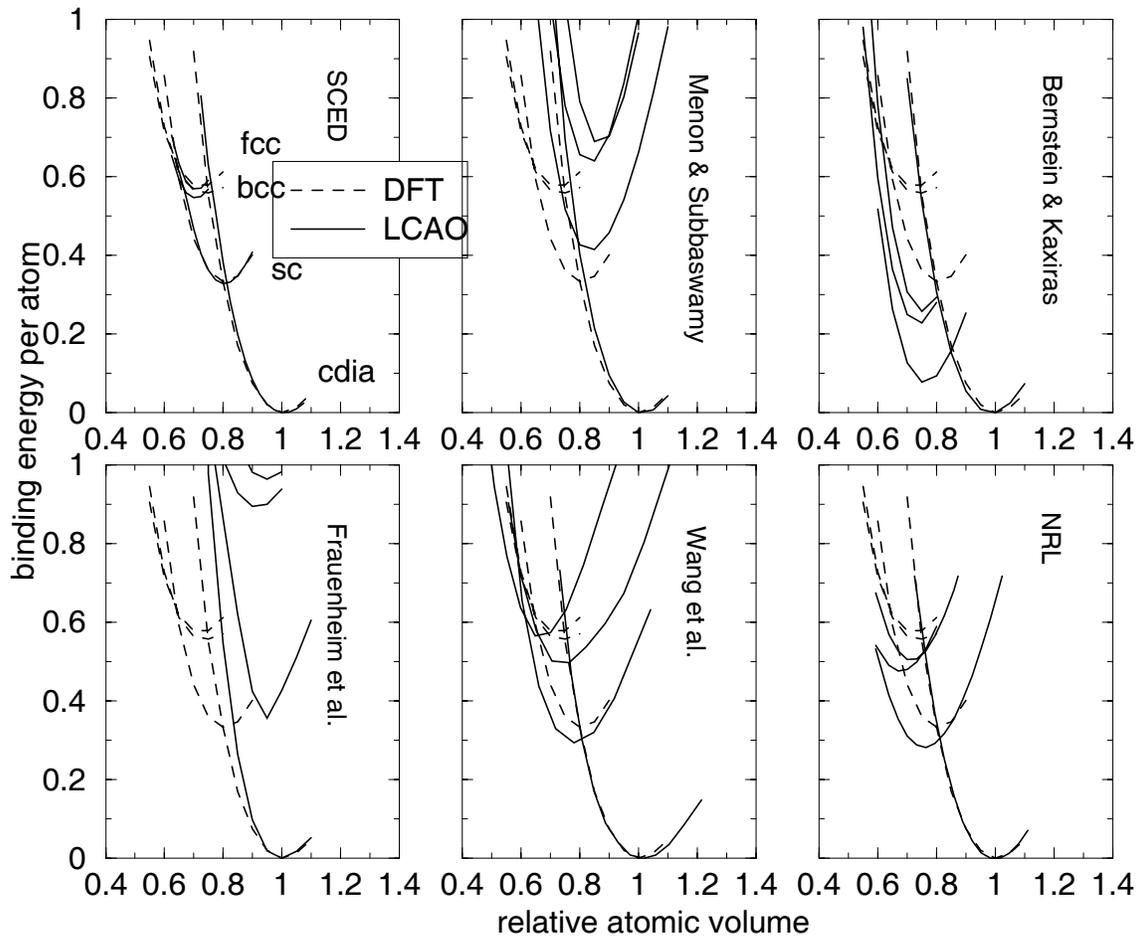

**Fig. 1**

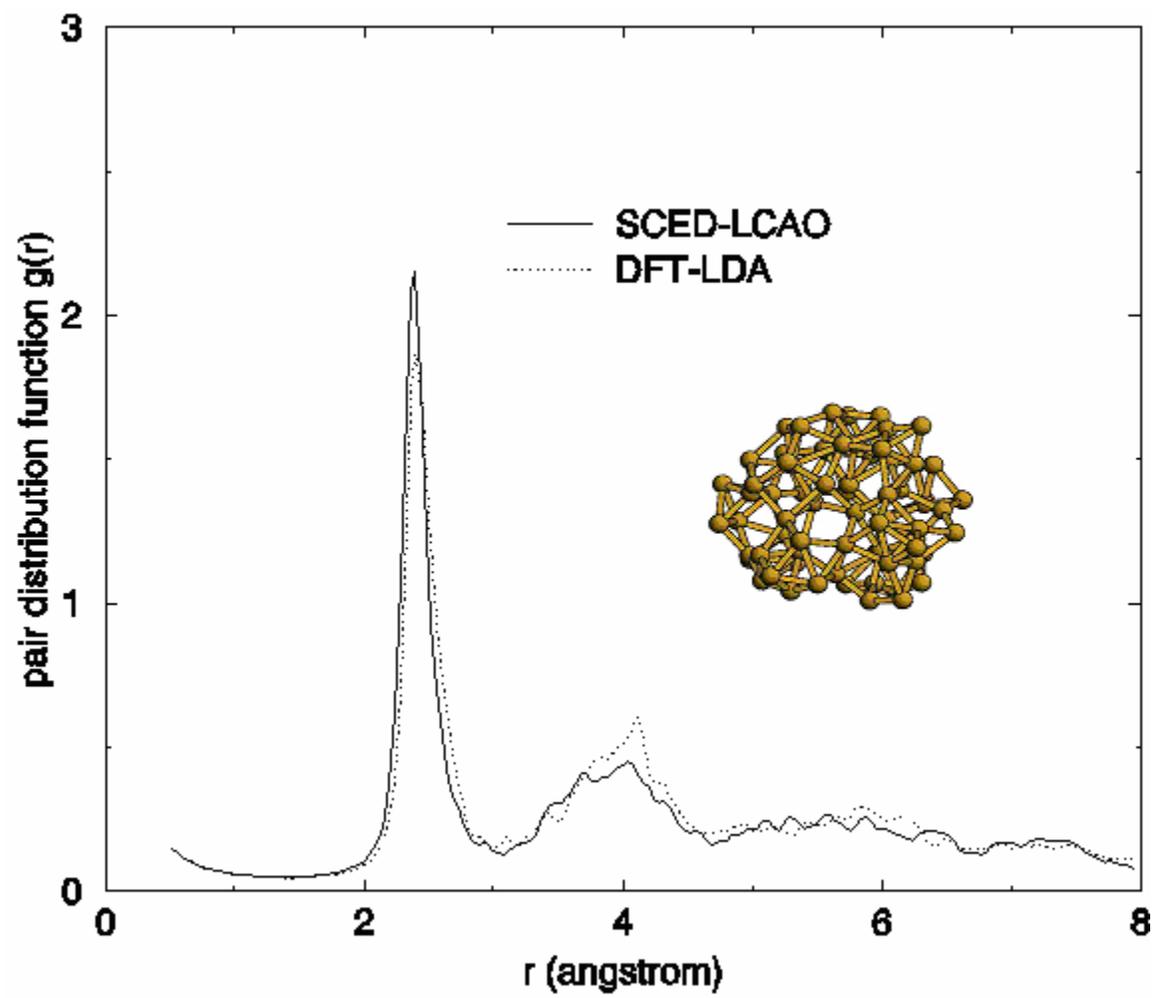

**Fig. 2**

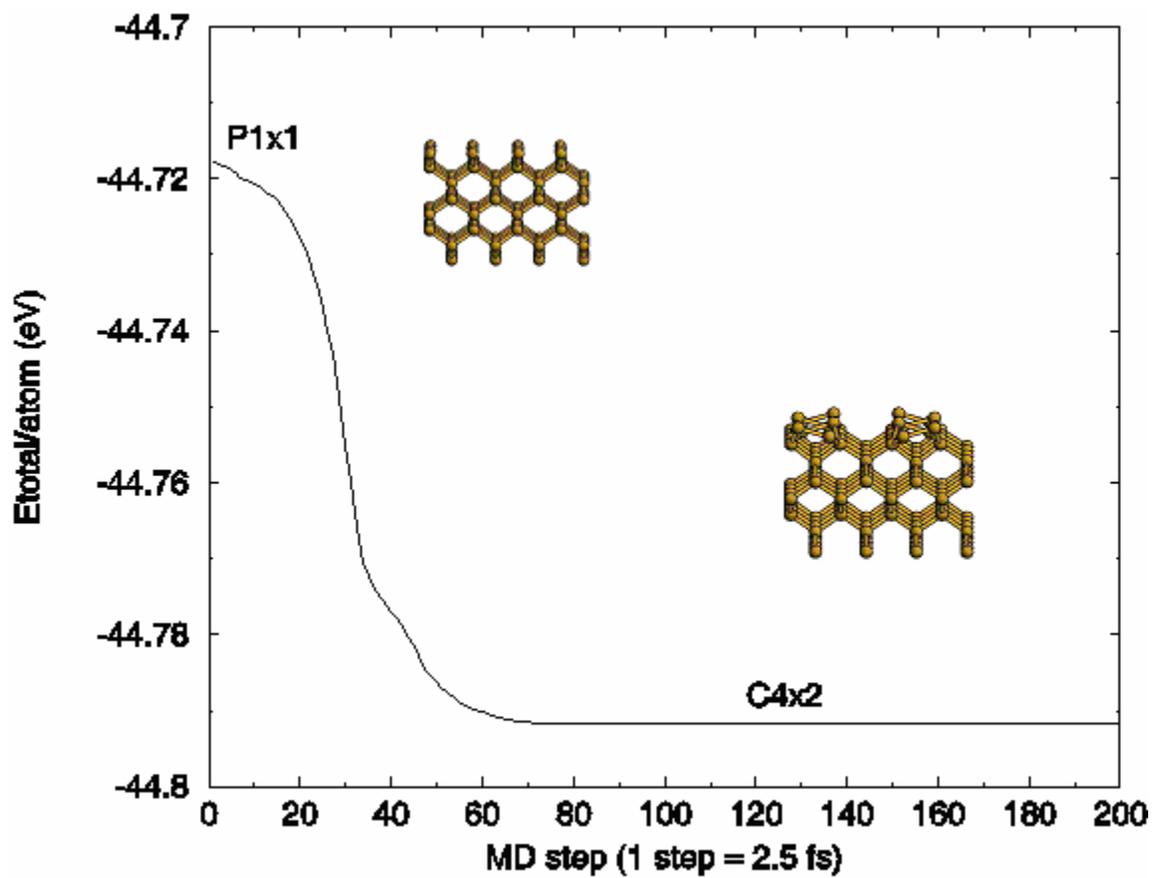

**Fig. 3**

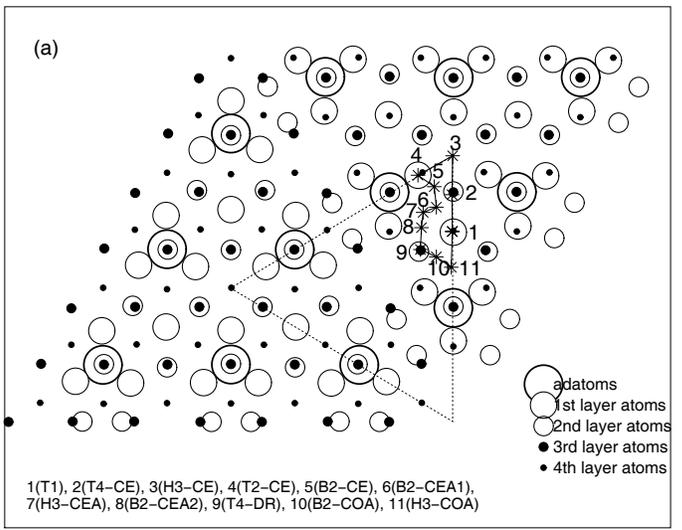

(a)

1(T1), 2(T4–CE), 3(H3–CE), 4(T2–CE), 5(B2–CE), 6(B2–CEA1),
7(H3–CEA), 8(B2–CEA2), 9(T4–DR), 10(B2–COA), 11(H3–COA)

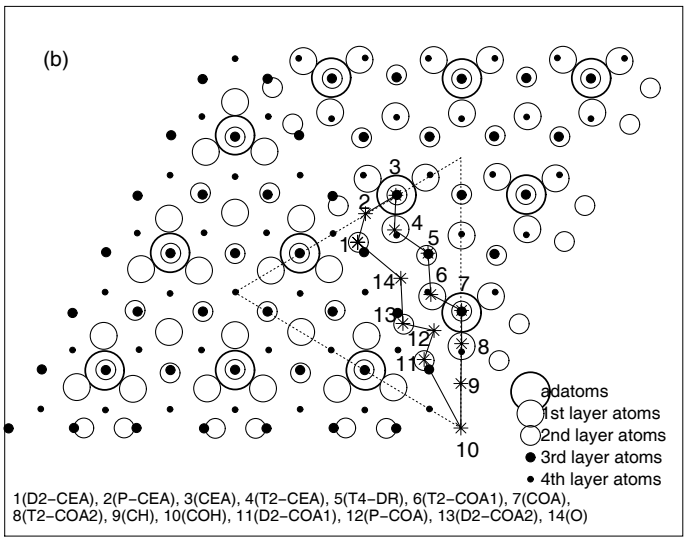

(b)

1(D2–CEA), 2(P–CEA), 3(CEA), 4(T2–CEA), 5(T4–DR), 6(T2–COA1), 7(COA),
8(T2–COA2), 9(CH), 10(COH), 11(D2–COA1), 12(P–COA), 13(D2–COA2), 14(O)

**Fig. 4**

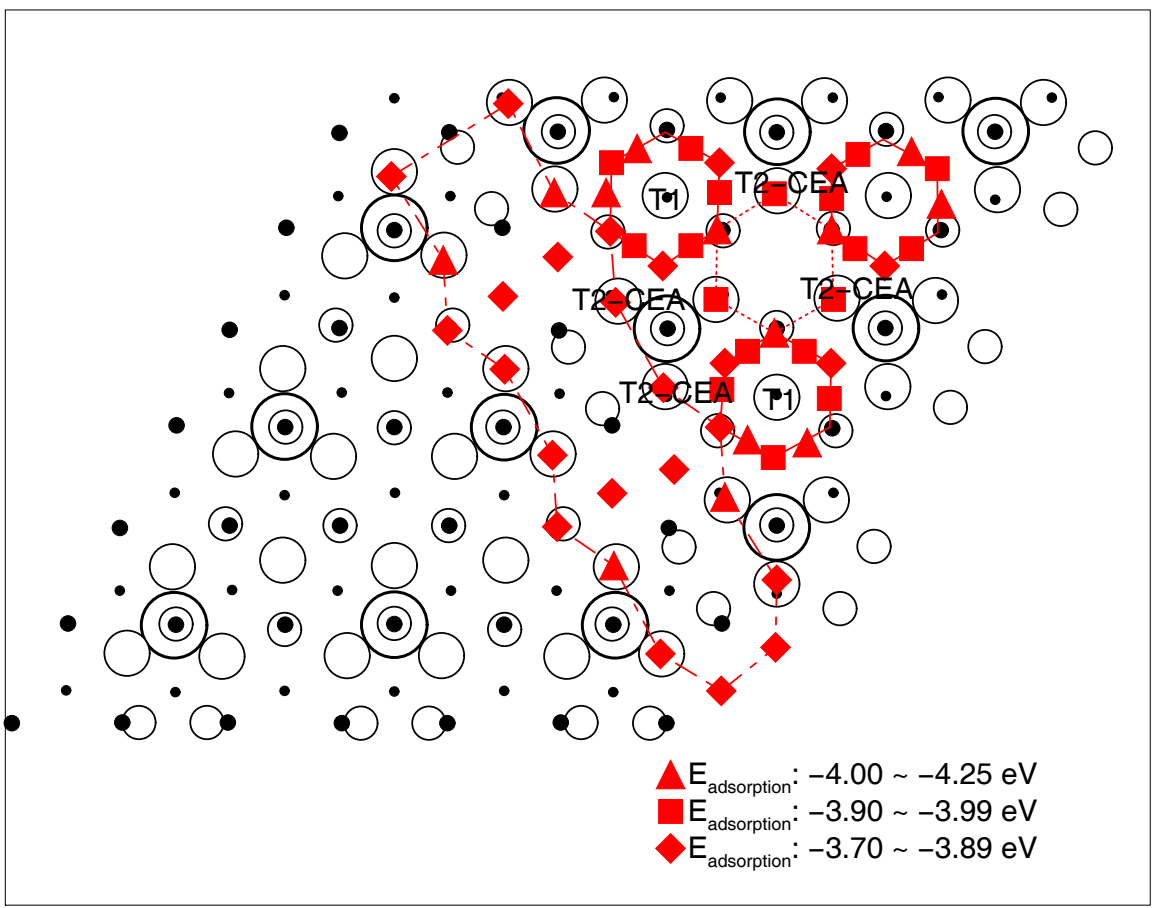

Fig. 5